\documentclass[twocolumn,prd,showpacs,nofootinbib]{revtex4}
\usepackage{epsf,graphicx}
\usepackage{amsmath,amsfonts,amssymb,latexsym}
\def\nq{\hspace*{-1em}}
\def\nqq{\hspace*{-2em}}

\def\cm{\hspace*{1cm}}
\def\inch{\hspace*{1in}}

\def\noi{\noindent}

\def\Jl#1#2{#1 {\bf #2},\ }

\def\ApJ#1 {\Jl{Astroph. J.}{#1}}
\def\CQG#1 {\Jl{Class. Quantum Grav.}{#1}}
\def\DAN#1 {\Jl{Dokl. AN SSSR}{#1}}
\def\GC#1 {\Jl{Grav. Cosmol.}{#1}}
\def\GRG#1 {\Jl{Gen. Rel. Grav.}{#1}}
\def\JETF#1 {\Jl{Zh. Eksp. Teor. Fiz.}{#1}}
\def\JETP#1 {\Jl{Sov. Phys. JETP}{#1}}
\def\JHEP#1 {\Jl{JHEP}{#1}}
\def\JMP#1 {\Jl{J. Math. Phys.}{#1}}
\def\NPB#1 {\Jl{Nucl. Phys. B}{#1}}
\def\NP#1 {\Jl{Nucl. Phys.}{#1}}
\def\PLA#1 {\Jl{Phys. Lett. A}{#1}}
\def\PLB#1 {\Jl{Phys. Lett. B}{#1}}
\def\PRD#1 {\Jl{Phys. Rev. D}{#1}}
\def\PRL#1 {\Jl{Phys. Rev. Lett.}{#1}}

\def\al{&}
\def\lal{&&{}}
\def\eq{Eq.\,}
\def\eqs{Eqs.\,}
\def\beq{\begin{equation}}
\def\eeq{\end{equation}}
\def\bear{\begin{eqnarray}}
\def\bearr{\begin{eqnarray} \lal}
\def\ear{\end{eqnarray}}
\def\earn{\nonumber \end{eqnarray}}

\def\nnn{\nonumber\\ \lal }
\def\nnnv{\nonumber\\[2pt] \lal }
\def\yy{\\[2pt] {}}

\def\eql{\al =\al}

\def\dst{\displaystyle}
\def\tst{\textstyle}
\def\fracd#1#2{{\dst\frac{#1}{#2}}}
\def\fract#1#2{{\tst\frac{#1}{#2}}}
\def\Half{{\fracd{1}{2}}}
\def\half{{\fract{1}{2}}}

\def\e{{\,\rm e}}
\def\d{\partial}

\def\const{{\rm const}}
\def\eps{\varepsilon}

\def\mn{_{\mu\nu}}

\def\MN{^{\mu\nu}}

\def\rh{r_{\rm hor}}
\def\xh{x_{\rm hor}}

\def\N{{\mathbb N}}

\def\asflat{asymptotically flat}
\def\sph{spherically symmetric}
\def\ssph{static, spherically symmetric}

\def\RN{Reiss\-ner-Nord\-str\"om}

\begin{document}
\thispagestyle{empty}
%\bigskip

\title{Dilaton gravity, charged dust, and (quasi-) black holes}

\author{K.A. Bronnikov}
\affiliation{Center for Gravitation and Fundamental Metrology, VNIIMS, 46 Ozyornaya
Street, Moscow 119361, Russia, and\\ Institute of Gravitation and Cosmology,
PFUR, 6 Miklukho-Maklaya Street, Moscow 117198, Russia, and\\
I. Kant Baltic Federal University, Alexander Nevsky Street 14, Kaliningrad 236041, Russia}
\email{kb20@yandex.ru}

\author{J.C. Fabris and \fbox{R. Silveira}}
\affiliation{Universidade Federal do Esp\'{\i}rito Santo, Departamento de F\'{\i}sica,
Av. Fernando Ferrari 514, Campus de Goiabeiras, CEP 29075-910, Vit\'oria, ES, Brazil}
\email{fabris@pq.cnpq.br}

\author{O.B. Zaslavskii}
\affiliation{Department of Physics and Technology, Kharkov V.N. Karazin
	National University, 4 Svoboda Square, Kharkov, 61077, Ukraine, and\\
	Institute of Mathematics and Mechanics, Kazan Federal University, 
	18 Kremlyovskaya Street, Kazan 420008, Russia}
\email{zaslav@ukr.net}

\begin{abstract}
	We consider Einstein-Maxwell-dilaton gravity with charged dust and
	interaction of the form $P(\chi) F\mn F\MN$, where $P(\chi)$ is an
	arbitrary function of the dilaton field $\chi$ that can be normal
	or phantom. For any regular $P(\chi)$, static configurations are
	possible with arbitrary functions $g_{00} = \exp(2\gamma(x^i))$
	($i=1,2,3$) and $\chi = \chi(\gamma)$, without any assumption of
	spatial symmetry. The classical Majumdar-Papapetrou system is
	restored by putting $\chi = \const$. Among possible solutions are
	black-hole (BH) and quasi-black-hole (QBH) ones. Some general
	results on BH and QBH properties are deduced and confirmed by
	examples. It is found, in particular, that \asflat\ BHs and QBHs can
	exist with positive energy densities of matter and both scalar and
	electromagnetic fields.
\end{abstract}

 %%({\bf check!})
\maketitle

%%\section{Introduction}

  An important type of static charged dust configurations is represented by
  the Majumdar-Papapetrou (MP) solution \cite{majumdar,papa}; it comprises
  an equilibrium between gravitational attraction and electric repulsion
  without any spatial symmetry assumption: equilibrium is
  established for any spatial shape of the charged dust cloud provided the
  charge to mass density ratio takes everywhere the proper value,
  $\rho_{e}/\rho_{m}=\pm 1$ in natural units ($c = G = 1$),

  The MP system was recently revived in a new context, that of the
  so-called quasi-black holes (QBHs) \cite{lemos1}--\cite{mein11}. Using
  the fact that in this solution the force balance implies a charge-to-mass
  ratio similar to that in the vacuum extremal Reissner-Nordstrom solution,
  a configuration has been proposed where such a starlike object has a size
  very close to the horizon radius. Such a system looks, for a
  distant external observer, quite similar to a true BH, though an
  event horizon has not been formed.

  We here extend this treatment to include a dilatonic scalar field, which
  can be partly motivated by studies in string theory. Along with general
  observations on possible equilibrium configurations [to be called
  dilatonic MP (DMP) systems), we consider BHs and QBHs supported by
  certain electric and scalar charge distributions. In particular, we try to
  find phantom-free configurations, i.e., those able to exist with
  positive-definite energy densities of matter and both fields.

  This problem has been considered in a PhD thesis of one of the co-authors
  of this paper, Robson Silveira, who died in 2009 before completing his
  study. He obtained some initial results indicating that such scalar
  QBHs are really possible and described some of their main properties.
  Our goal here is to briefly report on a more general analysis strongly
  developing his findings. A more detailed presentation can be found in
  Ref.\,\cite{we-13}.

%%%%%%%%%%%%%%%%%%%%%%%%%%%%%
%%\section {Basic equations}

  Consider the Lagrangian ($c=G=1$)
\beq    \label{L}
      L = \frac{1}{16\pi}\Big[
      		R + 2\eps (\d\chi)^2 - F^2 P(\chi)\Big]
					+ L_m + A_\mu j^\mu + J\chi,
\eeq
  where $\eps = \pm 1$ ($\eps=1$ for a normal scalar field $\chi$),
  $L_m$ is the Lagrangian of matter,
  $J$ is the scalar charge density,
  $F^2\equiv F^{\alpha\beta}F_{\alpha\beta}$
  ($F\mn = \d_\mu A_\nu - \d_\nu A_\mu$, the electromagnetic field),
  $j^\mu = \rho_e u^\mu$ is the 4-current, and $u^\mu$ is the 4-velocity.
  We do not fix the sign of $P(\chi)$ to provide correspondence with
  \cite{clem1,clem2}.  Following the ideas of the MP solution, we consider a
  static equilibrium with the metric
\bear                                                       \label{ds}
      ds^2 = \e^{2\gamma} dt^2 - \e^{-2\gamma} h_{ik} dx^i dx^k,
\ear
  and assume only the electric components $F_{0i} = -F_{i0} = \phi_i$ to be
  nonzero among $F\mn$; $\gamma$, $h_{ik}$, $\phi$, $\chi$ are functions of
  $x^i$, $i = 1,2,3$; $h_{ik}$ is the Euclidean flat metric, in general, in
  curvilinear coordinates. We use the notations $\gamma_i = \d_i\gamma$,
  $\phi_i =\d_i \phi$ etc; spatial indices are raised and lowered with the
  metric $h_{ik}$ and its inverse $h^{ik}$.  Also, $u^\mu = \delta^\mu_0
  \e^{-\gamma}$.

  The equations for $\chi$ and $\phi$ and the relevant combinations
  of the Einstein equations can be written in the following form:
\bear                                                   \label{eq-s1}
      	2\eps \e^{2\gamma} \Delta \chi
			+ P_\chi \phi_i \phi^i \eql - 8\pi J,
\yy                                                     \label{eq-F1}
         \nabla_i \left (\e^{-2\gamma}P \phi^i\right)
	 		 	\eql 4\pi \rho_e \e^{-3\gamma},
\yy                                                     \label{int}
	\e^{2\gamma} (\gamma_i \gamma^k + \eps \chi_i \chi^k)
			\eql P\, \phi_i \phi^k,
\yy							\label{rho_m}
	\e^{2\gamma} (\Delta\gamma - \gamma^i\gamma_i
		-\eps\chi^i\chi_i) \eql  4\pi \rho_m,
\ear
  where $\nabla_i$ and the Laplace operator $\Delta = \nabla_i \nabla^i$
  are defined in terms of the metric $h_{ik}$.
  \eq (\ref{int}) does not contain the densities, hence it holds both in
  vacuum and in matter; \eq (\ref{rho_m}) is a convenient expression for
  $\rho_m$ in terms of $\gamma(x)$ and $\chi(x)$. The Einstein
  equations also lead to the equilibrium condition
\beq                                                          \label{equi}
	  \rho_m \gamma_i - \rho_e \phi_i \e^{-\gamma} = J\chi_i.
\eeq

  The tensor equation (\ref{int}) implies that $\gamma$, $\chi$ and $\phi$
  are functionally related, and if $\gamma \ne \const$, we can put $\phi =
  \phi(\gamma)$, $\chi = \chi(\gamma)$; \eq (\ref{int}) then reduces to
\beq                                                        \label{int00}
	\e^{2\gamma} (1 + \eps \chi_\gamma^2) = P \phi_\gamma^2.
\eeq
  Hence we have the following arbitrariness: for any $P(\chi)$ and any 3D
  profile $\gamma(x^i)$, even more than that, for an arbitrary scalar field
  distribution $\chi = \chi(\gamma)$, we find $\phi(\gamma)$ from
  (\ref{int}), and the remaining field equations (\ref{eq-s1}),
  (\ref{eq-F1}) and (\ref{rho_m}) give us the mass, electric and scalar
  charge distributions that support this field configuration.

  In what follows we will try to obtain examples of BH and QBH
  configurations in the simplest case of spherical symmetry, and of special
  interest can be those where all kinds of matter are
  ``normal'', i.e., $ P > 0$, $\eps = +1$ and $\rho_m \geq 0$.

  The classical MP system is reproduced if we put $\chi = \const$,
  $P(\chi)\equiv 1$, and we necessarily obtain $|\rho_e| = \rho_m$. On the
  contrary, putting $\phi = \const$, we obtain MP-like systems with an
  arbitrary function $\gamma(x^i)$, existing only with a phantom $\chi$
  field, as follows from \eq (\ref{int00}).

%%%%%%%%%%%%%%%%%%%%%%%%%%%%%%%%%%%%%%%%%%%%%%%%%%%%%%%%%%%%%%%%%%%%%%%%
%%\section{Spherical symmetry: General features and examples}

  In the case of {\bf spherical symmetry}, the metric (\ref{ds}) reads
\beq                                                          \label{ds1}
	ds^2 = \e^{2\gamma} dt^2 - \e^{-2\gamma}(dx^2 + x^2 d\Omega^2),
\eeq
  where $x$ is a radial coordinate and $d\Omega^2$ is the line element on a
  unit sphere. The usual spherical (areal) radius is $r(x) = x \e^{-\gamma}$.
  Our set of equations takes the form
\bear                                                         \label{chi1}
	2\eps x^{-2}\e^{2\gamma}(x^2\chi')'
			+ P_\chi \phi'{}^2 \eql - 8\pi J(x),
\\                                                           \label{phi1}
	x^{-2} \big(P \e^{-2\gamma} x^2 \phi'\big)'
			\eql 4\pi \rho_e \e^{-3\gamma},
\\                                                           \label{rho_m1}
        \e^{2\gamma} (\gamma'' + 2\gamma'/x - \gamma'{}^2
		-\eps\chi'{}^2) \eql 4\pi \rho_m,
\\                                                           \label{int1}
	\gamma'{}^2 + \eps \chi'{}^2 \eql \e^{-2\gamma} P \phi'{}^2,
\\                                                           \label{equi1}
	\rho_m \gamma' - \rho_e \phi' \e^{-\gamma} \eql J\chi',
\ear
  where the prime denotes $d/dx$. The above arbitrariness transforms here
  into the freedom of choosing the functions $\gamma(x)$ and $\chi(x)$
  even if the coupling function $P(\chi)$ has been prescribed from the
  outset. All other quantities are then found from \eqs
  (\ref{chi1})--(\ref{equi1}).

  It is of interest how to choose the arbitrary functions in order to obtain
  a starlike configuration with a regular center or a BH. It is also of
  interest to seek phantom-free configurations such that $\eps =  +1$ and
  $\rho_m \geq 0$.

  A {\bf regular center} is obtained in the metric (\ref{ds1}) at $x=0$
  if and only if $\gamma(x) = \gamma_c + O(x^2), \ \gamma_c = \const$.
  Using a Taylor expansion for $\e^{2\gamma} \equiv A(x)$ at small $x$,
  one can show that $\rho_m > 0$ near the center requires that
  $g_{00} = A(x)$ should have there a minimum.

  Near a {\bf horizon} we must have $\e^{2\gamma}\sim (x-\xh)^n$, where
  $n \in \N$ is the order of the horizon. From (\ref{ds1}) it is clear that
  a horizon of finite radius $\rh = x\e^{-\gamma}\big|_{x=\xh}$ is only
  possible with $\xh = 0$ and $n =2$ (a {\it double, or extremal horizon}).
  Thus at small $x$ we can write  $A(x) =\half A_2 x^2 + \fract{1}{6} A_3 x^3
  + \cdots$, $A_i = \const, \ A_2 > 0$. Assuming  that $\chi$ and $\chi'$
  are finite at the horizon, we obtain $\rho_m \sim x^2$, but it can be of
  any sign without a direct correlation with $\eps$. From the field
  equations it follows that $\rho_e \sim x$ or possibly $\rho_e = o(x)$,
  while $J$ generically tends there to a finite limit. Thus such
  configurations, being in general perfectly regular and smooth, still
  contain an anomaly: the density ratios $\rho_e/\rho_m$ and $J/\rho_m$ are
  infinite at the horizon.

  For dust balls of finite size placed in vacuum, the external domain is
  described by the corresponding ``vacuum'' Einstein-Maxweel-dilaton (EMD) 
  solution; however, such
  solutions to the field equations are only known for some special choices
  of $P(\chi)$, e.g., $P = \e^{2\lambda\chi}$ \cite{BSh-77, dbh1, dbh2}.
  Therefore, instead, we consider {\bf \asflat} matter distributions with a
  smoothly decaying density. At large $x$ we can take
\beq                                                         \label{A-as}
       A(x) = 1 - \frac{2m}{x} + \frac{q_*^2}{x^2} + \cdots,
       \ \ \ \chi(x) = \chi_\infty + \frac{\chi_1}{x} + \cdots,
\eeq
  and \eq (\ref{rho_m1}) then yields
\beq                                                         \label{rho_ma}
       4\pi \rho_m = \frac{1}{x^4} (-3m^2 + q_*^2 - \eps \chi_1^2)
       			+ o(x^{-4}).
\eeq
  This clearly shows that large charges $q$ are necessary for obtaining
  $\rho_m > 0$ if $\eps = +1$. (Note that the extreme \RN\ solution
  with the charge $q =m$ corresponds in the notation
  (\ref{A-as}) to $q_*^2 = 3m^2$.)
  The densities $\rho_e$ and $J$ also behave in general as $1/x^4$ at
  large $x$.

{\bf Integral charges.} The field at flat spatial infinity is characterized
  by integral charges: the electric charge $q$ such that the electric field
  strength is $\phi' = q/x^2 + o(1/x^2)$, the scalar charge $D$ such that
  $\chi' = D/x^2 + o(1/x^2)$, and the mass $m$ corresponding to the
  Schwarzschild asymptotic $\e^\gamma \approx 1-m/x$, hence $\gamma'
  \approx m/x^2$ (note that $x \approx r$ at large $x$).  A relation between
  these three quantities directly follows from \eq (\ref{int1}). Indeed,
  multiply (\ref{int1}) by $x^4$ and take the limit $x \to \infty$ to obtain
\beq
	 m^2 - q^2 + \eps D^2 =0,                      \label{charges}
\eeq
  since $\e^\gamma \to 1$ and $P\to 1$ (assuming that a weak electromagnetic
  field should be Maxwell). This generalizes a similar relation (2.12) from
  \cite{clem2}, written there for vacuum EMD systems with $P(\chi)
  \sim \e^{2\lambda\chi}$.

  Thus, as compared to the MP system where $q = \pm m$, a balance in the DMP
  system requires $m^2 > q^2$ if $\eps  = -1$ (both electric and phantom
  scalar fields are repulsive), but $m^2 < q^2$ with a canonical, attractive
  scalar field.

  \eq (\ref{charges}) is valid for all \asflat\  (islandlike) EMD
  systems since they are approximately \sph\ in the asymptotic region.

{\bf Quasi-black holes.}
  By definition, in some region $r\leq r^*(c)$ of a QBH it holds that
  $\e^{\gamma} \sim c$, where $c$ is a small parameter, and the limit
  $c\to 0$ usually corresponds to a BH. The most general \ssph\ QBH in our
  problem setting is a system with the metric (\ref{ds1}) and a regular
  center, and at small $x$ we can write
\beq
     \e^{2\gamma} \equiv A(x,c) = A_0 (c) + \half A_2 (c) x^2 + \cdots,
\eeq
  where $A_0 (c) \to 0$ as $c \to 0$ while $A_2(0)$ is finite.
  Without loss of generality we can assume
\beq
	\e^{2\gamma} = \frac{x^2 + c^2}{f^2(x,c)},        \label{gam-q}
\eeq
  where $f$ is a smooth function that has a well-defined nonzero limit
  $c\to 0$. The value $c=0$ in (\ref{gam-q}) corresponds to an extreme BH
  metric with a horizon at $x=0$. In particular, taking $f(x,0) = x+m$, we
  obtain the extreme \RN\ metric. At  small enough $c$ and $x\lesssim
  c$, $e^{2\gamma} = O(c^2)$ is arbitrarily small.

  Let us stress that, given (\ref{gam-q}), the region  where the ``redshift
  function'' $\e^\gamma$ is small, is itself not small at all. Indeed,
  suppose $f(x,c) = O(1)$, and $c \ll 1$. Then the radius $r(c)$ of the
  sphere $x=c$ (which belongs to the high redshift region) is
  $f(c,c)/\sqrt{2} = O(1)$; the distance from the center to this sphere,
  $\int_0^c \e^{-\gamma}\,dx$, is also $O(1)$.

\medskip\noi    %%%%%%%%%%%%%%%%%%%%%%%%%%%%%%%%%%%%%%%%%%%%%%%%%%%
{\bf Example 1.} Let us choose the metric function
\beq
	\e^\gamma = \frac{z}{m+ 2z -y},                        \label{A4}
	\ \  y := \sqrt{x^2 + a^2},
	\ \  z := \sqrt{x^2 + c^2},
\eeq
  with certain positive constants $m,\ a,\ c$. At small and large $x$ we
  have
\bear \nqq                                                   \label{A4_0}
	x\to 0: \nq && \e^{2\gamma} = \frac{c^2}{(m{-}a{+}2c)^2}
			+ x^2\,\frac{m {-} a {+} c^2/a}{(m{-}a{+}2c)^3}
\nnn \inch
			+ O(x^4),
\yy   \nq
	x\to \infty: \nq && \e^{2\gamma} = 1 - \frac{2m}{x}  \label{A4_i}
		 + \frac{3m^2 {+} a^2 {-} c^2}{x^2} + O(x^{-3}).
\ear
  The system has a regular center and is \asflat, and $m$ is
  the Schwarzschild mass. Assuming
\beq
	c < a < m,                                           \label{par4}
\eeq
  we can be sure that $\rho_m > 0$ near the center since $\e^\gamma$ has a
  minimum there (see above). For $\rho_m$ there is a bulky expression
  leading to $\rho_m > 0$ for proper choices of the dilaton field profile
  $\chi(x)$ with $\eps=+1$ under the condition (\ref{par4}).
%% fig 1 ----------------------------------
%\begin{figure}
%\centering
%\includegraphics[width=6.2cm]{fig2c-q2.pdf}
%\caption {The function $\rho_m(x)$ in Example 1 for $m=1$, $a=0.5$,
%	$c = 0.1$ and $b = 0.39,\ 0.395,\ 0.398,\ 0.4$ in a range of $x$
%	where the $b$ dependence is significant (upside down).	  }
%\end{figure}
%% ----------------------------------
  It is the case, for instance, if we assume
\beq                                                        \label{chi4}
	\chi' = b/y^2, \cm b = \const > 0
\eeq
  with sufficiently small $b$.
%  Fig.\,1 shows that, with the chosen values of
%  $m,\,a,\,c$, we obtain $\rho_m > 0$ if $b < b_0 \approx 0.398$.

  The expressions for the electric and scalar charge densities are bulky,
  but their particular form can add nothing to our understanding of the
  situation; it is only important that they are finite and regular.

  The limit $c \to 0$ leads to an extreme BH metric,
\beq
	\e^\gamma = \frac{x}{m+ 2x -y}, \cm                     \label{A3}
  	y := \sqrt{x^2 + a^2}.
\eeq
  We thus obtain an \asflat\ BH without phantoms. With (\ref{chi4}) for
  $\chi$ and $\eps = +1$, we obtain from (\ref{A3})
\beq                                                         \label{rho_m3}
	4\pi \rho_m = \frac{x^2[(a^2+b^2)y - b^2(2x+m)]}{y^4 (2x-y+m)^3}.
\eeq
  We have $\rho_m > 0$ at all $x > 0$ in a certain region of the parameter
  space. Thus, putting $m=1$ (fixing the units) and $a=0.5$ (for example),
  we find that $\rho_m > 0$ for $0 < b < b_0 \approx 0.369$.

  The expressions for $\rho_e$ and $J$ are cumbersome; it is only important
  that, for a generic choice of $P(\chi)$, they are everywhere finite and
  regular and behave at the horizon as described above.

\medskip\noi
{\bf Example 2.} Our framework allows for describing polycentric systems,
  with any number of mass concentrations. For instance, one can consider
  the metric (\ref{ds}) in Cartesian coordinates $x^i=(x,y,z)$ (so that
  $h_{ik}= \delta_{ik}$) and choose
\beq                                                         \label{f_n}
	\e^{-\gamma(x^i)}\equiv  f(x^i)
		= \frac{1}{n} \sum_{a=1}^{n} f_a (X_a),
\eeq
  where $f_a$ are functions of $X_a := |x^i - x^i_a|$, $x^i_a$ being the
  (fixed) coordinates of the $a$-th center. As $f_a$, one can take any
  functions providing \asflat\ \sph\ solutions, e.g., BHs or QBHs.
  A complete solution is obtained after choosing the function
  $\chi(\gamma)$, or equivalently $\chi(f)$, which should be regular at all
  relevant values of $f$ and decay sufficiently rapidly at spatial infinity,
  as $f \to 1$.

% fig 3 ----------------------------------
\begin{figure*}
\centering
\includegraphics[width=5.7cm]{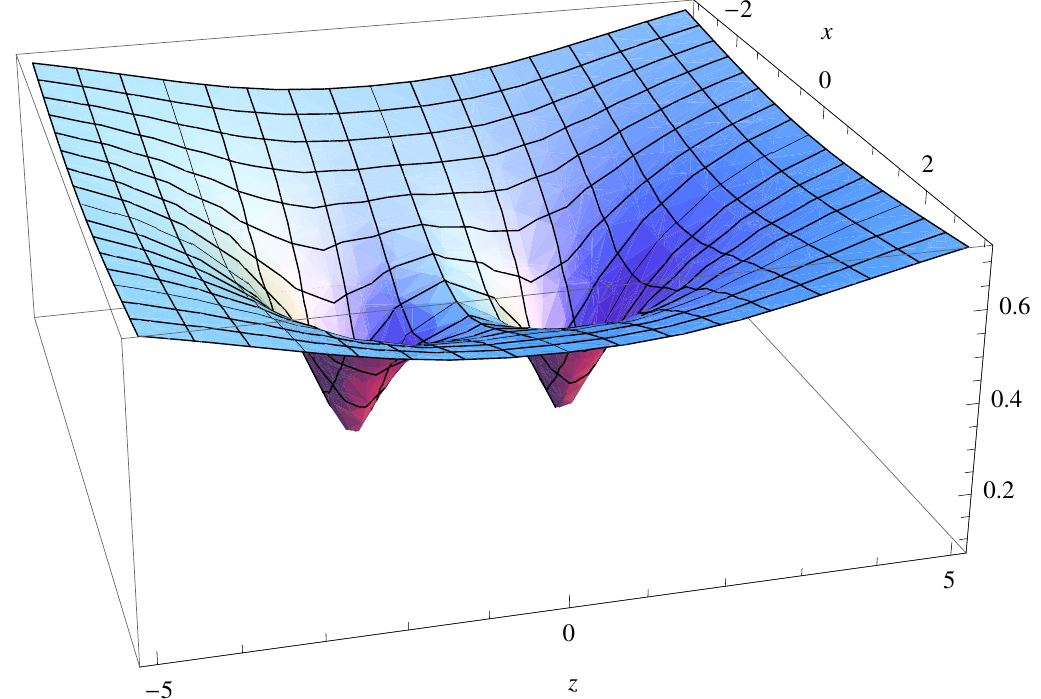}\
\includegraphics[width=5.7cm]{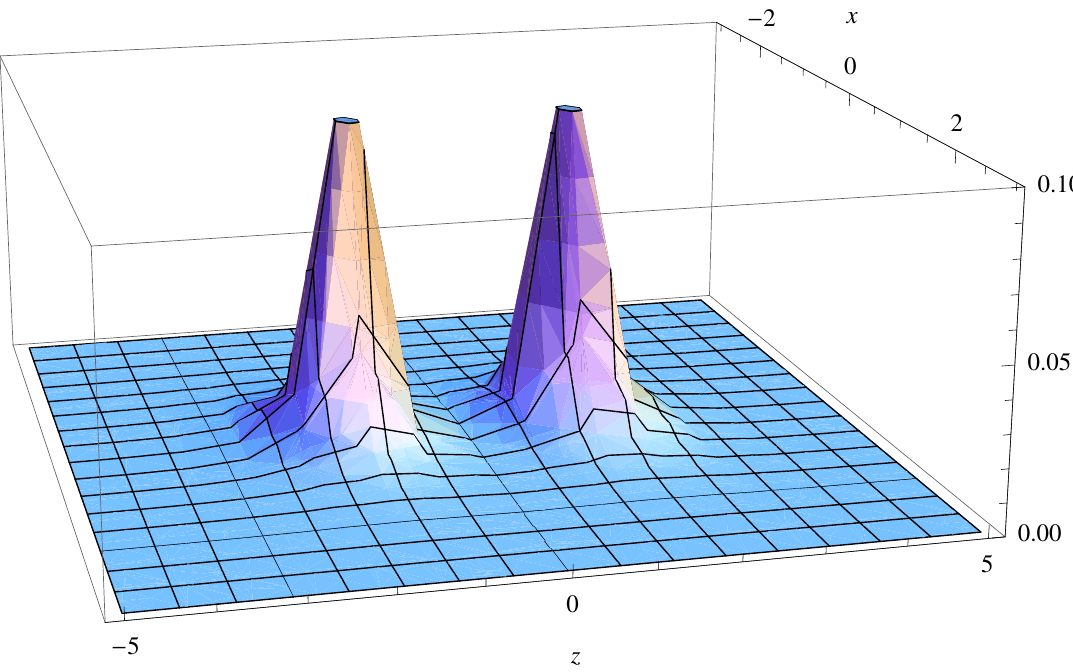}\
\includegraphics[width=5.7cm]{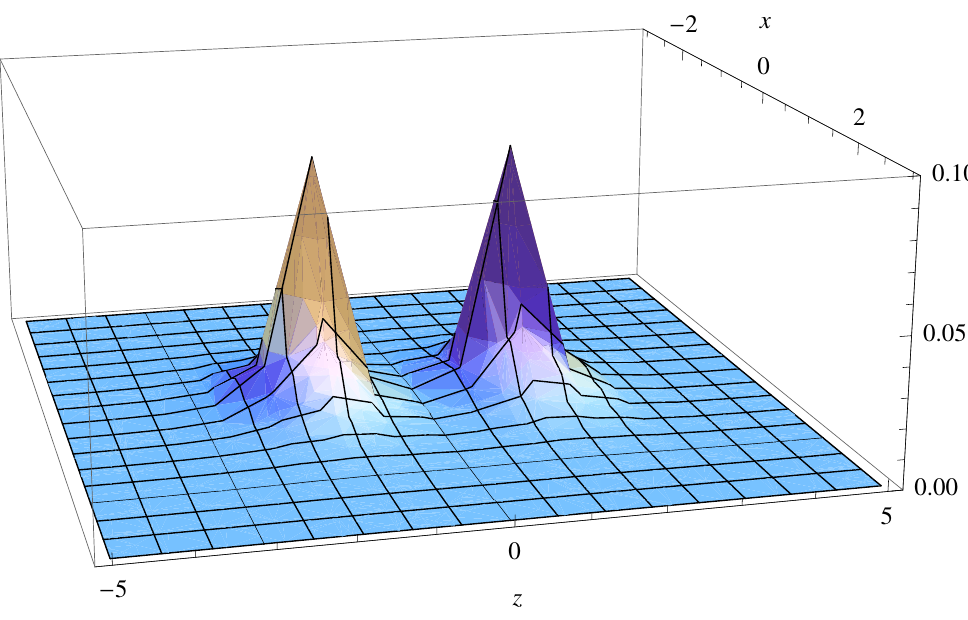}
\caption{Plots for Example 2, sections $y=0$ of different 3D profiles for a
	system of two identical QBHs. {\bf Left:} the metric function
	$\e^{2\gamma (x,y,z)}$ for $m_1=m_2=1$, $a=1.5$, $c_1=c_2 = 0.2$.
	{\bf Middle:} the density $\rho_m(x,y,z)$ for the same parameters
	and $b=0$, i.e., for a pure MP system.  {\bf Right:} the same for
	$\eps=+1$ and $b = 0.07$, i.e., for a DMP system with the specified
	$\chi(f)$.  }
\end{figure*}
%     ---------------------------------

  What follows is an example of a system of two QBHs: let
\bearr                                                        \label{f_3}
      f(x^i) = \frac{m_1+z_1}{2 z_1} + \frac{m_2+z_2}{2 z_2},
\\ \lal                                                    \label{chi_3}
      \chi(f) = \Half b(f-1)^2,
\\ \lal
      z_1 := \big(|\vec x - \vec x_1|^2 + c_1^2 \big)^{1/2},
\ \ \
      z_2 := \big(|\vec x - \vec x_2|^2 + c_2^2 \big)^{1/2},
\nnnv
      \vec x_1 = (0,\ 0,\ a), \ \ \ \ \vec x_2 = (0,\ 0,\ -a),
\earn
  with constants $m_1 > 0,\ m_2 > 0,\ a > 0,\ b \geq 0$, $c_1\geq 0$ and
  $c_2\geq 0$. The electric potential $\phi$ and all densities are found
  from \eqs (\ref{int00}), (\ref{eq-s1}), (\ref{eq-F1}), and (\ref{rho_m}).
  In particular, for the mass density we obtain
\bearr
	4\pi\rho_m = \frac{1}{f^2(x^i)} \biggl[
		\frac{3m_1 c_1^2}{z_1^2 (m_1+z_1)^3}
\nnn \cm
		+ \frac{3m_2 c_12^2}{z_2^2 (m_2+z_2)^3}
		- \eps b^2 (f-1)^2 f^i f_i \biggr].
\ear

  The special case $b=0$ corresponds to a bicentric MP configuration.
  If $c_1$ or $c_2$ is zero, the corresponding ``center'' is a
  BH, while at small nonzero $c_a$ it is a QBH.

  Figure 1 shows the 3D behavior of the metric function $e^{2\gamma} \equiv
  f^{-2}(x^i)$ and the mass density $\rho_m (x^i)$ for the chosen example of
  a system of two QBHs for the specified parameter values. Evidently, the
  density is everywhere positive in both cases in Fig.\,1 [middle (a MP system)
  and right (a DMP system with a canonical scalar field)], although inclusion of
  a scalar field makes it smaller.

  In conclusion, let us enumerate the main results.

1. It has been shown that, with the Lagrangian (\ref{L}), static
   configurations are possible with arbitrary functions $g_{00} =
   \e^{2\gamma(x^i)}$ ($i=1,2,3$) and $\chi = \chi(\gamma)$, for any regular
   coupling function $P(\chi)$, without any assumption of spatial symmetry.

2. There are purely scalar analogs of MP systems, but only with
   phantom scalar fields.

3. There is a universal balance condition, (\ref{charges}), between the
   Schwarzschild mass and the electric and scalar charges, valid for
   any \asflat\ DMP systems, including those with horizons and/or
   singularities. It generalizes the results previously obtained for
   special cases (e.g., \cite{clem2}).

4. In the case of spherical symmetry, the existence conditions have
   been formulated for BH and QBH configurations with smooth
   matter, electric charge and scalar charge density distributions.
   It turns out that horizons in DMP systems are second-order (extremal),
   in agreement with the general properties of QBHs \cite{lemos4}.

5. Examples of phantom-free \sph\ BH and QBH solutions have been obtained,
   and an example of a phantom-free system of two QBHs.
\medskip

\subsection*{Acknowledgments}

  We thank CNPq (Brazil) and FAPES (Brazil) for partial financial support.

\end{document}